# Controllable Thermo-Stimulated Luminescence in Niobate Persistent Phosphor by Constructing the Photovoltaic/Electrolytic Cell for Remote Intelligent Anti-Counterfeiting

Yuanyuan Hu[a], Dangli Gao[a,*], Xiangyu Zhang[b], Sining Yun[c]

[a] *School of Environmental and Municipal Engineering, Xi'an University of Architecture and Technology, Xi'an, Shaanxi 710055, China*

[b] *College of Science, Chang'an University, Xi'an, Shaanxi 710064, China*

[c] *Functional Materials Laboratory (FML), School of Materials Science & Engineering, Xi'an University of Architecture and Technology, Xi'an, Shaanxi 710055, China*

*Corresponding Author

E-mail addresses: gaodangli@163.com, gaodangli@xauat.edu.cn

## ABSTRACT

Persistent luminescence (PersL) carrying remote key information plays a crucial role for intelligent anti-counterfeiting applications. However, the weak PersL intensity accompanied by uncontrollability limits their practical application. Here we develop $LiNbO_3$ (LNO):Pr,Bi phosphor with enhanced red PersL by trace doping $Sm^{3+}$. The LNO:Pr,Bi,Sm phosphor exhibits quadruplet luminescence, including polychrome photoluminescence, PersL, and photo/thermo-stimulated luminescence (PSL/TSL). Particularly, the enhanced TSL can carry remote subjective information independent of the phosphor itself by controlling the temperature. A mechanism of afterglow enhancement is proposed based on constructing reversible photovoltaic cells and electrolytic cells by photothermal redox reactions using $Bi^{3+}$ + $V_O$•• and $Bi^{3+}/Pr^{3+}$ + $V_{Li}'$ ion pair. This study has sparked the exploration of designing the information storage PersL materials for more sophisticated remote intelligent anti-counterfeiting.

KEYWORDS: persistent phosphor, deep-trap persistent luminescence, intelligent anti-counterfeiting, trace doping

## 1. Introduction

Persistent luminescence (PersL) material, that can storage and release charge carriers for energy storage and conversion, is a passive photovoltaic material. PersL materials have great potential for various applications such as display, biological imaging, optoelectronic communication, sensing, anti-counterfeiting, and information storage due to its low energy consumption, fast response, and visual multi-color and multi-mode display [1-5]. Especially, PersL materials with expanding the capacity of the visualized information are frequently used for advanced anti-counterfeiting and various imaging devices [6-10]. However, these luminescent materials still lack memory function, which limits their further application in remote intelligent anti-counterfeiting field.

The PersL of luminescent storage materials can last a few minutes to hours after cessation of light irradiation. As is well known, PersL comes from the binding between the afterglow center and the captured electrons in the traps [11-15]. These traps, that are various types of defects in materials, can trap the produced holes or electrons, leading to storage and memory functions [16,17].

The PSL/TSL, that is different from PersL, results from a combination between luminescent center and the carriers released by traps under photon/thermo-stimulation in a pre-charging phosphor [18,19]. Therefore, the "writing" and "reading" of information are recorded through the bistable state of traps with/without carriers [20,21]. Based on this principle, Wang et al. achieved information storage and reading in the $Ca_3Ga_4O_9$:$Bi^{3+}$ with quasi-layered structure using UV and NIR light, respectively [22].

Similarly, information access was achieved in activated $ZnAl_{1.4}Ge_{0.3}O_{3.7}:Cr^{3+}$ PersL phosphor by dual beams of UV and NIR light [23]. However, due to the weak PersL intensity with uncontrollable nature, current afterglow anti-counterfeiting is an objective event dependent on luminescent characteristics of phosphors. For intelligent anti-counterfeiting, materials need to have the ability to remember human-computer interaction information, which poses a new challenge on luminescent anti-counterfeiting materials.

The PersL material with memory function is expected to be used for intelligent anti-counterfeiting. Such PersL material must meet the following requirements: 1) Good resistance to environmental heat disturbance and large signal-to-noise ratio [24,25]; 2) High carrier storage capacity; 3) Strong stimulated luminescence intensity that can be controlled as needed for effective information read-out ability. Inorganic PersL phosphor is one of the best candidates for information storage due to its excellent thermal stability, erasability, reversibility, and tunable luminescent properties [26-28]. However, most of the inorganic PersL materials currently studied typically exhibit weak PSL/TSL, which is difficult to achieve remote information storage. Therefore, developing strong and controllable PSL/TSL materials remains a huge challenge for achieving remote information storage and anti-counterfeiting.

Lithium niobate ($LiNbO_3$, abbreviated as LNO) is a multi-functional material with excellent piezoelectric/ferroelectric, electro-optic, and nonlinear optical properties, and is one of the best candidates in display, luminescent printing anti-counterfeiting, radio frequency filters, optical information storage, communications, and other fields.

Previous studies have shown that doping rare earth (RE) ions can generate not only photoluminescence (PL) and photochromism [29], but also mechanical luminescence (ML) [30,31]. Besides doping RE, doping non-rare earth ions such as Bi, Fe, Mn, V, Mg, and Zn in LNO can greatly enhance its ability to response to light or resist photorefractive damage [32-36]. Many advances and applications have been also made in RE or/and $Bi^{3+}$ doping LNO luminescent materials [37-39]. However, obtaining multi-mode fluorescence in LNO with memory function still remains a challenge. The core reason lies in the lack of effective control over the trap and energy level structure of multi-mode fluorescent materials. The currently recognized and effective method for regulating traps and energy level positions is doping strategy, that can tune the position of energy levels and enhance trap density by matrix distortion and introducing new energy levels. The Sm and the main group metal element Bi both have variable valence and multiple coordination structures. Considering that it is not easy to replace $Nb^{5+}$ in the lattice with $Bi^{3+}$ in LNO matrix due to non-equivalent substitution, we attempted to distort the local lattice through trace-doping $Sm^{3+}$, opening a gap for $Bi^{3+}$ doping to enter the matrix lattice. Then, based on the domino effect, $Bi^{3+}$ successfully replaced $Nb^{5+}$ into matrix lattice. Slightly doping $Sm^{3+}$ is to avoid the introduction of impurity energy levels of doped ions, which may cause non-radiative relaxation of the luminescent center.

Based on the above analysis, here, we intentionally designed and developed LNO:Pr,Bi phosphor with trace $Sm^{3+}$ doping that shows excellent multi-modal fluorescence including PL, PersL, PSL, photo-stimulated PersL (PSPL) and TSL. As expected, the stimulated luminescence can be enhanced by trace-doping $Sm^{3+}$. The multi-mode luminescence mechanism is also suggested based on Bi-O covalent bond regulation [36]. As expected, the phosphors demonstrate unique dynamic fluorescent pattern encrypted by contrasting visualized temperature distribution, which holds a promise in intelligence anti-counterfeiting and temperature visualization sensors fields [40-47].

## 2. Experimental section

### *2.1 Materials*

Chemicals: The compounds of high-purity $LiCO_3$ (99.99%), $Nb_2O_5$ (99.95%), $Bi_2O_3$ (99.99%) $Pr_6O_{11}$ (99.999%) and $Sm_2O_3$ (99.99%) are all derived from Aladdin. Polyvinyl alcohol (PVA, GR) was purchased from China National Pharmaceutical Group Chemical Reagent Company.

### *2.2 Sample synthesis*

The phosphors with the chemical formula $Li_{1-x-z}Nb_{1-y}O_3$:$xBi^{3+}$,$yPr^{3+}$,$zSm^{3+}$ (x=0, 0.003, 0.008, 0.01 and 0.015; y=0, 0.002, 0.005, 0.008 and 0.01; z=0, $1\times10^{-10}$, $1\times10^{-8}$, $1\times10^{-6}$ and $1\times10^{-4}$) were developed via solid-state reaction approach. Raw materials are weighed by chemical stoichiometry, and place them in an agate mortar, mechanically grinding for 30 min to obtain a sufficiently fine mixture. Afterwards, we transfer the obtained mixture to an alumina crucible in a furnace, and then the mixture is calcined at a heating rate of 5 °C $min^{-1}$ in air at 1100 °C for 9 h. Finally, collect naturally cooled products and grind them into fine powder again for performance characterization.

*2.3 Printing the luminescent images*

PDMS-based resin and curing agent are mixed in a certain proportion and then the mixture was mechanically stirred for 6 min. Subsequently, we poured the as-prepared phosphors into the mixture and thoroughly stirred for achieving luminescent inks. Finally, the anti-counterfeiting pattern was printed on non-fluorescent card using inks.

*2.4 Characterization*

A D/Max2550GBt/PC X-ray diffractometer (XRD) was employed to perform crystal phase detection on the synthesized sample under 1.5418 Å Cu Kα (40 kV, 40 mA) irradiation. A scanning electron microscope (SEM, Zeiss Gemini 300) with energy dispersive X-ray spectroscopy (EDX) was used to analyze the morphology, size, and chemical composition of the product. The spectroscopic properties of the sample are characterized using a fluorescence spectrophotometer (Horiba PTI Quanta Master 8000) with a 150 W xenon lamp, a 5 W 365 nm ultraviolet lamp, and power adjustable 808 nm and 980 nm near-infrared lasers as irradiation sources. TL curves were acquired by using a custom-made heating apparatus integrated into spectrofluorometer (Horiba PTI) of (temperature range, 25-500 °C; heating rate, 1 °C $s^{-1}$). Laser diodes were programmed to scan the luminescent pattern and the photographs of pattern were captured using a Cannon camera (EOS 60D) under the assistance of appropriate optical filters.

**3. Results and discussion**

*3.1 Structure and Morphology*

The XRD diffraction peaks of all samples are consistent with the standard card (PDF # 85-2456) of $LiNbO_3$ in Fig. 1a, indicating that the dopant has successfully entered the LNO lattice. The Rietveld structure refinements (Fig. 1b) demonstrate $LiNbO_3$ belongs to a trigonal crystal system with R3c space group and lacks inversion symmetry (Fig. 1c). According to the Hume Rothery rule, doped $Bi^{3+}$ (r = 1.03 Å, CN = 6) and $Pr^{3+}$ (r = 0.99 Å, CN = 6) ions are expected to substitute for $Li^+$ (r = 0.76 Å, CN = 6) ions at octahedral sites ($LiO_6$, top inset of Fig. 1c), while $Bi^{5+}$ (r = 0.76 Å, CN = 6) is more inclined to occupy the sites of $Nb^{5+}$ (r=0.64 Å, CN=6) [48-50]. Typically, $Bi^{5+}$ with same charge occupied the $Nb^{5+}$ sites will not cause the formation of defects. While $Bi^{3+}$/$Pr^{3+}$ occupying $Li^+$ sites, two Li vacancies ($V_{Li}'$) are generated simultaneously for charge balance. The defects created in the LNO structure could act as traps, generating the PersL, PSL, TSL, and ML. The SEM pattern in Fig. 1d shows that all samples, including LNO:1%Bi, LNO:1%Bi,0.5%Pr and LNO:1%Bi,0.5%Pr,0.0001%Sm, exhibit polyhedral shapes with sizes ranging from 5~10 μm, indicating that the dopant has little effect on the morphology of the samples. The EDS elemental mapping images of LNO:1%Bi,0.5%Pr,0.0001%Sm sample show that Li, Nb, O, Bi, and Pr are uniformly distributed throughout the lattice, indicating that Bi and Pr are uniformly doped into LNO. Since the $Sm^{3+}$ in the sample has an extremely small concentration, it cannot be observed.

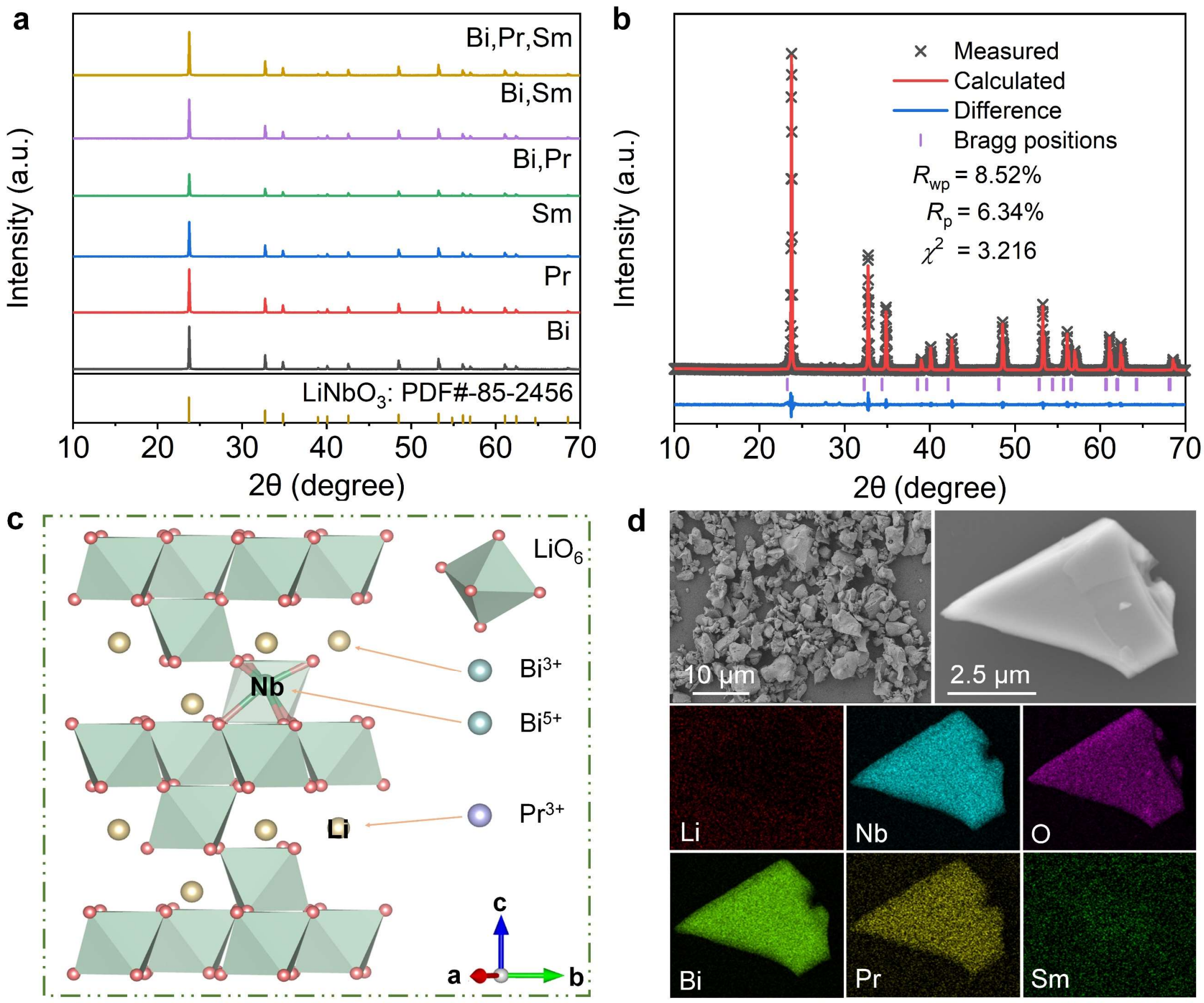

**Fig. 1.** Characterization of phase and structure of LNO-base phosphors. (a) XRD patterns. (b) Rietveld refinement of LNO:1%Bi,0.5%Pr,0.0001%Sm phosphor. (c) Crystal structure and possible lattice sites occupied by doping elements. (d) SEM and elemental mapping images of the selected LNO:1%Bi,0.5%Pr,0.0001%Sm particles.

*3.2 Spectral Analysis*

As is well known, the luminescence characteristics strongly depend on the concentration of dopant. The dopant concentration-dependent thermoluminescence (TL) curves are shown in Fig. S1. Based on the TL curves, the optimal doping concentration of $Pr^{3+}$, $Bi^{3+}$ and $Sm^{3+}$ in LNO:Bi,Pr,Sm are 0.5%, 1.0% and 0.0001%. It is worth noting that the doping concentration (0.0001%) of $Sm^{3+}$ is different from the conventional dopant concentration. $Sm^{3+}$ of 0.0001% is less than the impurity ion concentration in the dopant $Pr_6O_{11}$ with a purity of 99.999%. Just like the famous butterfly effect, trace amounts of $Sm^{3+}$ doping opened a domino effect, inducing the substitution of non-equivalent doped $Bi^{3+}$ for $Li^+$. In subsequent studies, the doping concentrations of $Pr^{3+}$, $Bi^{3+}$ and $Sm^{3+}$ were fixed at 0.5%, 1.0% and 0.0001%, respectively, i.e., LNO:1.0%Bi,0.5%Pr,0.0001%Sm.

Fig. 2a,b shows the comparisons of PL and PL excitation (PLE) spectra, respectively, for a series of samples including LNO:Bi,Pr,Sm, LNO:Bi,Pr, LNO:Bi,Sm, LNO:Bi, LNO:Pr and LNO:Sm. We divided the samples (excluding LNO:Bi) into two groups based on PL and PLE spectra. One set of samples containing Pr (i.e., LNO:Bi,Pr,Sm, LNO:Bi,Pr and LNO:Pr) mainly displayed $Pr^{3+}$ typical transitions, and another set of samples containing $Sm^{3+}$ (i.e., LNO:Bi,Sm and LNO:Sm) demonstrated $Sm^{3+}$ characteristic transitions. In LNO:Sm and LNO:Bi,Sm phosphors, a few PL peaks at 569 nm, 604 nm and 655 nm (Fig. 2a) can be easily assigned to the $^4G_{5/2}\rightarrow{}^6H_{5/2}$, $^4G_{5/2}\rightarrow{}^6H_{7/2}$ and $^4G_{5/2}\rightarrow{}^6H_{9/2}$ characteristic transitions of $Sm^{3+}$, respectively [51]. Monitoring $Sm^{3+}$ emission at 655 nm, PLE peaks at 287 nm and 408 nm are ascribed to Sm-O charge-transfer band (CTB) and $^6H_{5/2}\rightarrow{}^4L_{13/2}$ transition of $Sm^{3+}$, respectively [52]. While LNO:Bi,Pr,Sm, LNO:Bi,Pr, and LNO:Pr samples exhibit $^3P_0\rightarrow{}^3H_6$ characteristic emission of $Pr^{3+}$ (Fig. 2a) [15]. PLE broad bands in UV and blue regions can be ascribed to 4f→5d and $^3H_4\rightarrow{}^3P_0$ transitions of $Pr^{3+}$ in Fig. 2b [53].

Note that only LNO:Bi phosphor displays a broadband blue emission peaking at 400 nm, ascribing to the $Bi^{3+}$: $^3P_1 \rightarrow {}^1S_0$ transition (Fig. 2a) under irradiation of 246 nm UV light. While monitoring PL emission at 400 nm, PLE spectrum (Fig. 2b) exhibits a band peaking at 246 nm, assigning to the $^1S_0 \rightarrow {}^1P_1$ energy level transition of $Bi^{3+}$. Upon careful observation, all samples share a weak emission peak at 716 nm, assigning to the transition of the  O related defects [54]. Comparing the PLE and PL spectra, we can conclude that $Bi^{3+}$ only serve as a defect center in all co-doped samples due to the absence of $Bi^{3+}$ transitions in both their PL spectra and PLE spectra in all samples in Fig. 2a,b. In addition, based on the same absence of $Sm^{3+}$ transitions on PL and PLE spectra, we can also conclude that in LNO:Bi,Pr,Sm sample, $Sm^{3+}$ also only serves as defect centers, and then $Pr^{3+}$ serves as the only fluorescent center. As expected, the co-doping of $Bi^{3+}$ and $Sm^{3+}$ into LNO:Pr resulted in red fluorescence enhancement of more than ten times.

Besides PL, the three samples with $Pr^{3+}$ (i.e., LNO:Bi,Pr,Sm, LNO:Bi,Pr, and LNO:Pr) exhibit an afterglow peak at 620 nm accompanied with a weak peak at 716 nm, assigned to $^3P_0 \rightarrow {}^3H_6$ of $Pr^{3+}$ and matrix defects, respectively, in Fig. 2c. The PersL duration of all three samples is greater than 6 min, with the order being in LNO:Bi,Pr,Sm > LNO:Bi,Pr > LNO:Pr (Fig. 2d). The other three samples including LNO:Bi,Sm, LNO:Sm and LNO:Bi did not display in Fig. 2c, since they did not show PersL performance.

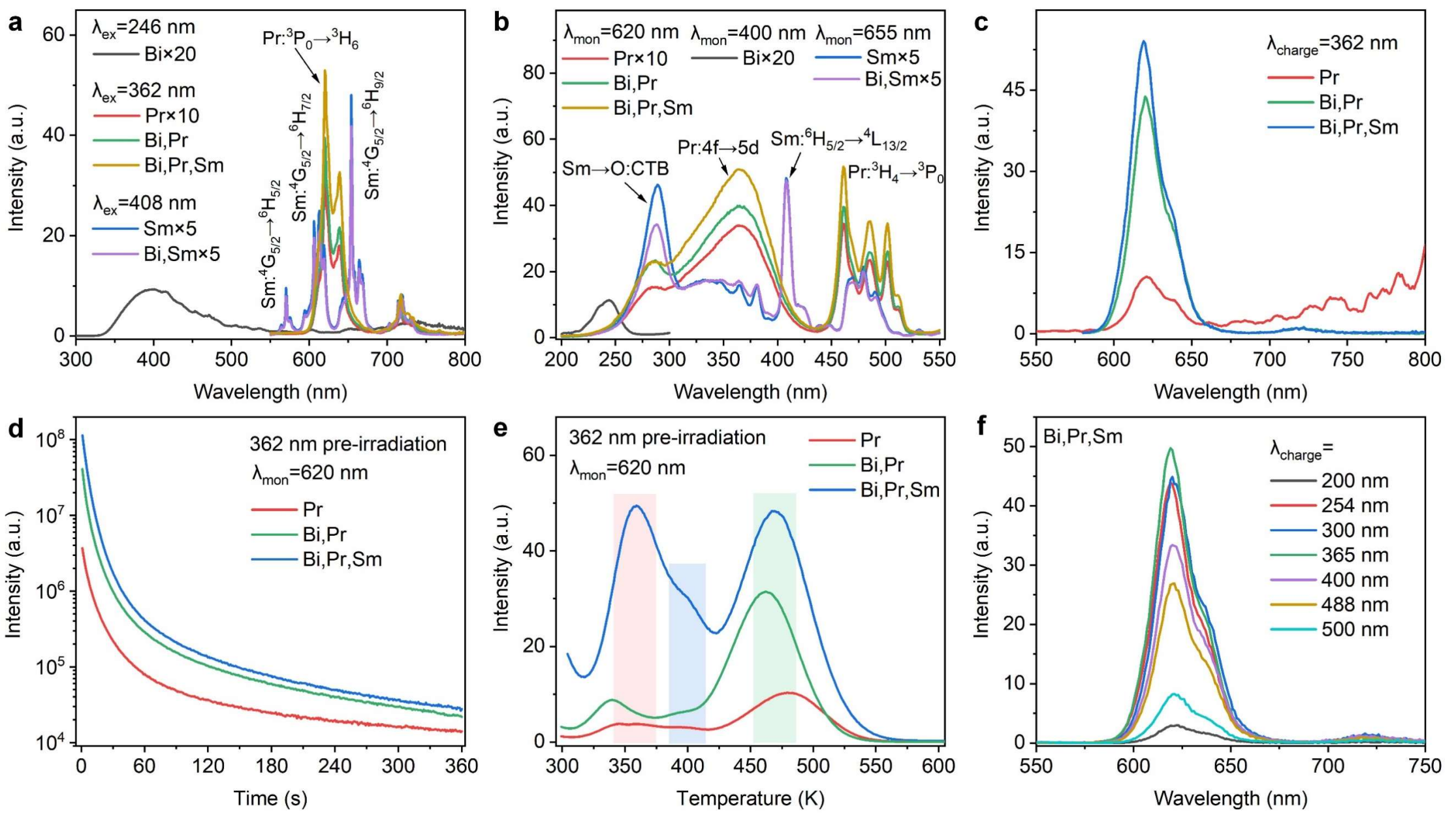

**Fig. 2.** Double-mode luminescence (PL and PersL) characteristics of LNO:Bi,Pr,Sm phosphor, with the spectra of LNO:Bi, LNO:Pr, LNO:Sm, LNO:Bi,Pr, and LNO:Bi,Sm phosphors as references. (a) PL spectra upon 246 nm, 362 nm, and 408 nm light excitation. (b) PLE spectra monitoring at 400 nm, 620 nm, and 655 nm. (c) PersL emission spectra recorded at 5 s decay. (d) PersL decay curves monitored emission at 620 nm after charging with 362 nm UV light for 5 min. (e) TL curves. (f) Charging wavelength-dependent PersL emission spectra. Before measuring the PersL, PersL decay curve, and TL, the phosphor was pre-irradiated with 362 nm UV light for 5 min.

Fig. 2e shows TL curves of the samples containing $Pr^{3+}$ (i.e., LNO:Bi,Pr,Sm, LNO:Bi,Pr, and LNO:Pr). TL curves in all samples containing $Pr^{3+}$ show the similar contours, composed of three peaks at about 359 K, 398 K and 468 K, respectively, except for different intensities (Fig. 2e). The TL integrated area represents trap density, while the peak represents trap depth. One can find that the LNO:Bi,Pr,Sm sample demonstrates great enhancement on trap density, especially on shallow trap peaking at 359 K, relative to LNO:Bi,Pr and LNO:Pr samples. We approximately estimated the trap depth E using the following equation [55]:

$$E = (0.94\ln\beta + 30.09) \times kT_m \quad (1)$$

where, $E$, $\beta$, and $k$ represent trap depth, heating rate, and Boltzmann constant, respectively, while Tm represents the temperature of the TL curve peak. According to the above equation, the trap depths of peak 1, 2, and 3 are approximately 0.93, 1.03 and 1.21 eV, respectively.

All the PersL emission spectra (Fig. 2f) from LNO:Bi,Pr,Sm are similar to its PL spectra (Fig. S2), with the two peaks at 620 and 716 nm, assigning to $Pr^{3+}$: $^3P_0 \rightarrow {}^3H_6$ transition and matrix defects. Unlike other afterglow materials that can be charged most effectively with UV photon, LNO:Bi,Pr,Sm can be pre-charged by UV and blue light.

*3.3 Trap Manipulation of Persistent Phosphors by Temperature Control*

To control PersL, Fig. 3 shows the luminescence signals under multiple fields stimulation. After removing the UV or visible light source, the afterglow gradually disappears with the delay of time. After that, two characteristic emission peaks assigned to $Pr^{3+}$: $^3P_0 \rightarrow {}^3H_6$ transition of and $O_i$ can be illuminated again under the irradiation of 808/980 nm or heating (Fig. 3). The position of the peaks on PSL, PSPL and TSL emission spectra agree well with their position on PersL spectra (Fig. S3) except for the relative intensity, confirming that the emission originated from the same fluorescent center. In addition, the pulse signal composed of PSL and PSPL can be further tuned by the stimulation of different periodic pulse near-infrared laser fields. As shown in Fig. 3a, the PSL intensity is correlated with the power and wavelength of near-infrared laser. After noticing several pulses, there was almost no significant decrease in PSL (Fig. 3a). In order to explain why the intensity of PSL remains basically unchanged after multiple cycles of light stimulation, we further tested the TL curve after 1 min of light stimulation. The TL curve demonstrated that a very small part of carriers in the shallow trap region were released under NIR light stimulation (Fig. 3b,c).

Besides PSL and PSPL (Fig. 3a), the PersL curve can also be thermally modulated, and the enhanced TSL can be obtained as needed by managing the heating temperature range (Fig. 3d,e) in LNO:Bi,Pr,Sm phosphor. Consequently, charge carriers stored in traps at different depths are conveniently extracted by addressing different depth traps under different heating energy stimulation. Compared to heating-controlled TSL of LNO:Bi,Pr (Fig. 3f), LNO:Bi,Pr,Sm phosphor demonstrated a greater advantage in TSL on-demand control due to the convenient control of the carrier releasing from traps with different depths (Fig. 3d-f). Comparing PSL with TSL in Fig. S3, it is easy to find that their relative intensity ratios of the two luminescent peaks are different. This may be due to the fact that photo-stimulation leads to not only PSL of O defects, but also up-conversion luminescence. Therefore, thermo-stimulation is a cleaner and more convenient way than NIR laser stimulation for control of afterglow on-demand of information extraction. Especially, the biggest advantage of thermal stimulation is that external stimuli do not tamper with stored information, while light stimulation often accompanies the writing of new information such as up-conversion during information extraction.

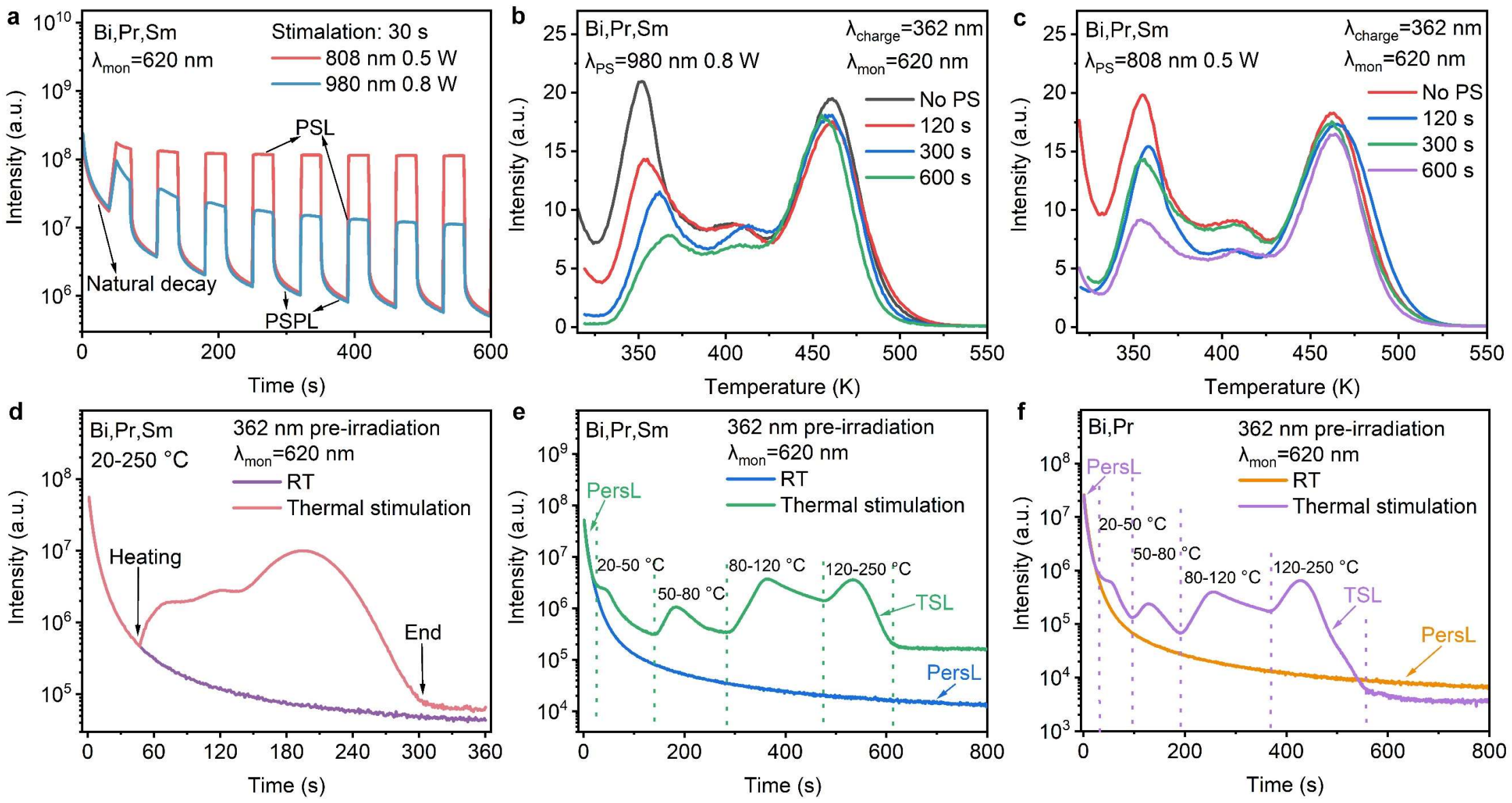

**Fig. 3.** Tuning the stimulation luminescence of LNO:Bi,Pr,Sm and LNO:Bi,Pr phosphors by light or heating on a natural decay curve. (a) Comb shaped luminescence signal composed of PSL and PSPL in a decaying LNO:Bi,Pr,Sm sample upon the irradiation of a pulse 980 nm (0.8 W)/808 nm (0.5 W) laser diode (on/off, 30/40 s). (b,c) TL curves monitoring at 620 nm after 980/808 nm NIR laser stimulation of pre-irradiated LNO:Bi,Pr,Sm. (d-f) TSL monitoring at 620 nm under thermal stimulation at varying temperature ranges of LNO:Bi,Pr,Sm (d,e) and LNO:Bi,Pr (f) phosphors. Before measuring the PersL decay curve, the phosphor was pre-charged with 362 nm UV light for 5 min.

To shed more light on the mechanism of enhanced PersL/TSL, we further investigated the temperature-dependent PersL characteristics of LNO:Bi,Pr,Sm phosphor in Fig. 4a-c, with the spectra of LNO:Bi,Pr (Fig. 4d-f) and LNO:Bi (Fig. S4) as references. These three phosphors can be divided into two groups: one group including LNO:Bi,Pr,Sm and LNO:Bi,Pr displays the temperature-dependent emission characteristic (Fig. 4); the other group (i.e., LNO:Bi phosphor) only displays the emission characteristic at room temperature (Fig. S4). One can observe that the emission spectra of LNO:Bi,Pr,Sm and LNO:Bi,Pr phosphors show the similar temperature-dependent relationship, featuring a decreases on the intensity of PersL with increasing temperature (Fig. 4a,d). The afterglow decay curves of the two samples also show that the afterglow intensity decreases with increasing temperature. However, it is interesting to find that at some special temperatures, the afterglow decay curve exhibits an extremely slow decay rate (Fig. 4b,e). By comparing the temperature-dependent TL curves, we found that these special temperatures happen to be at the peak temperature of the TL curve (Fig. 4c,f). As the temperature increases from 20 to 200 °C, shallow trap carriers are released under thermal loading, while deep trap carriers are sufficient to withstand the thermal disturbances (Fig. 4b,e). Note that temperature-dependent TL curves in Fig. 4c,f show that when heating sample to a certain temperature on the x-axis, the carriers in trap located at the low temperature side a certain heating temperature can be completely released. This provides a strategy for reading out information by addressing trap carrier information in a certain area on TL curves via controlling the temperature. Compared to PSL information read-out, it is a more effective way to address and read the stored information of traps in a certain depth interval (Fig. 3a). LNO:Bi,Pr,Sm demonstrates the optimum spectrum feature to adapt to high temperature working environment relative to LNO:Bi (Fig. S4).

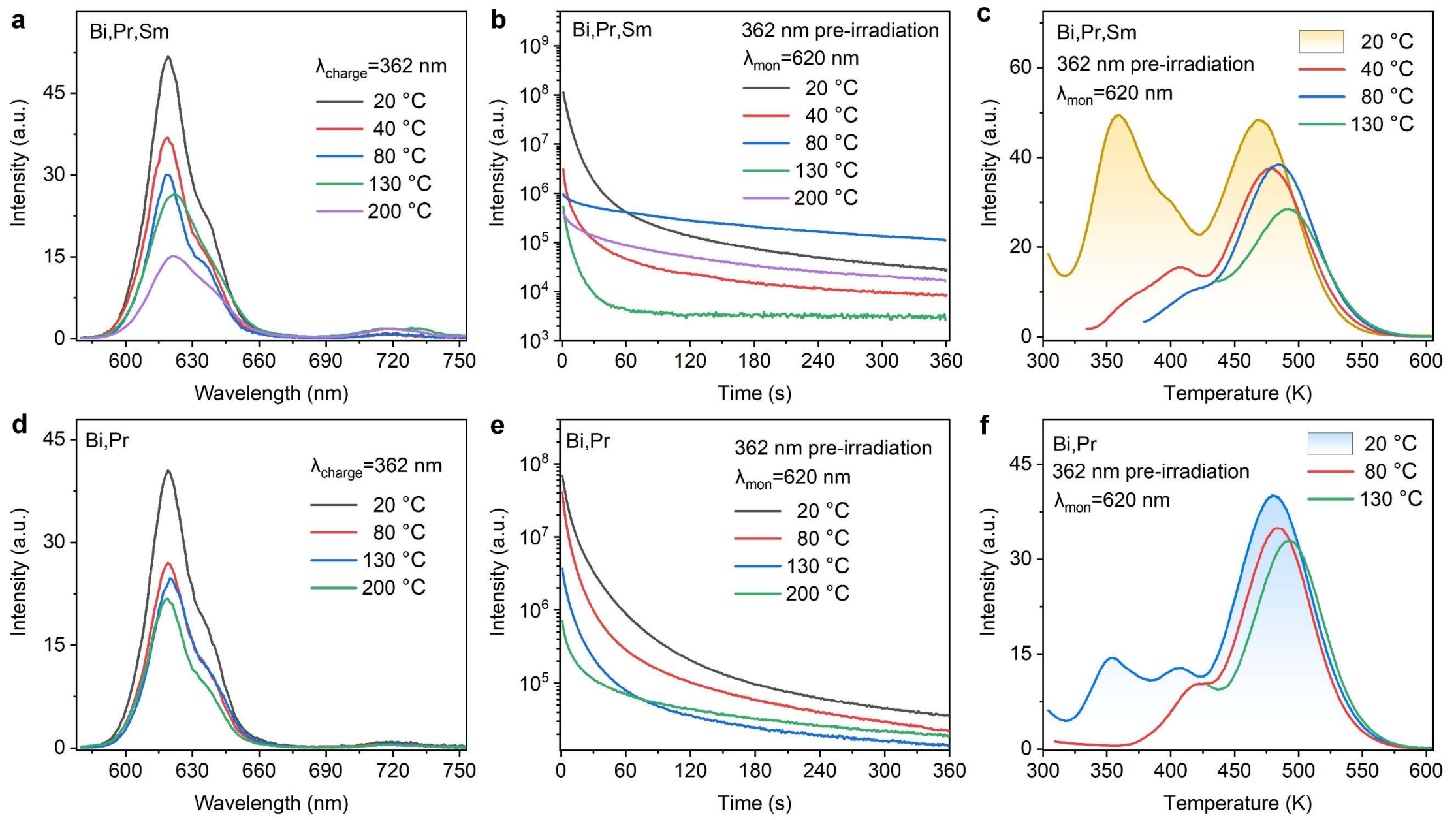

**Fig. 4.** Charging wavelength-dependent spectral characteristics of LNO:Bi,Pr,Sm and LNO:Bi,Pr phosphors. (a-c) PersL spectra, PersL decay curves, and TL curves of LNO:Bi,Pr,Sm. (d-f) PersL spectra, PersL decay curves, and TL curves of LNO:Bi,Pr. Prior to the measuring of PersL, PersL decay curves, and TL, the phosphor was pre-charged with 362 nm UV light for 5 min.

To reveal the processes of the trapping and releasing of charge carriers, Fig. 5 displays dynamic PL (monitoring at 620 nm from $Pr^{3+}$ transition) process of LNO:Pr, LNO:Bi,Pr and LNO:Bi,Pr,Sm phosphors before and after charging. When excited into luminescent center energy level using specific resonance wavelength, the PL dynamic process describes the competitive ability of trapping electrons from conduction band (CB) between luminescent center and trap center (Fig. 5). The shaded area in the Fig.

5 represents the trapping carriers by the trap from CB. The PL intensity monitoring 620 nm in LNO:Pr and LNO:Bi,Pr,Sm phosphors increase gradually with the prolonging time of 461 nm blue light irradiation, which means that traps have strong competitiveness in capturing of charge carriers and they should be in the vicinity of $Pr^{3+}$. Moreover, after charging with 254 nm UV light, the PL intensity did not decrease, indicating that even after UV charging, the number of $Pr^{3+}$ fluorescent center ions did not decrease and his role as the fluorescent center (Fig. 5a,c). However, under the same blue irradiation, the traps do not have good competitiveness relative to $Pr^{3+}$ due to the transient reaching saturation of PL in LNO:Bi,Pr in Fig. 5b, announced a remote defect distribution around $Pr^{3+}$. After 254 nm UV light charging, $Pr^{3+}$ did show a decreasing trend in PL, indicating that part $Pr^{3+}$ act as the trap. By monitoring the weak PL emission originating from matrix defects in the three samples (Fig. S5a,b,d) except for LNO:Bi,Pr (Fig. S5c), we found that as the charging time increased, the PL intensity decreased, confirming the essential role of matrix defects as traps. Based on these results, we conclude that trace doping $Sm^{3+}$ may perturb the $Bi^{3+}$ doped sites, allowing $Bi^{3+}$ occupy both $Li^{+}$ and $Nb^{5+}$ sites in LNO:Bi,Pr,Sm. As a result, $V_{Li}'$ and $V_O^{\bullet\bullet}$ defects are generated at $Bi_{Li}$ and $Bi_{Nb}$ octahedra, respectively, for the charge balance, which protects $Pr^{3+}$ as only a fluorescent center. While in LNO:Bi,Pr, $Bi_{Li}$ and $Pr_{Nb}$ octahedra stabilized $V_{Li}'$ and $V_O^{\bullet\bullet}$, respectively.

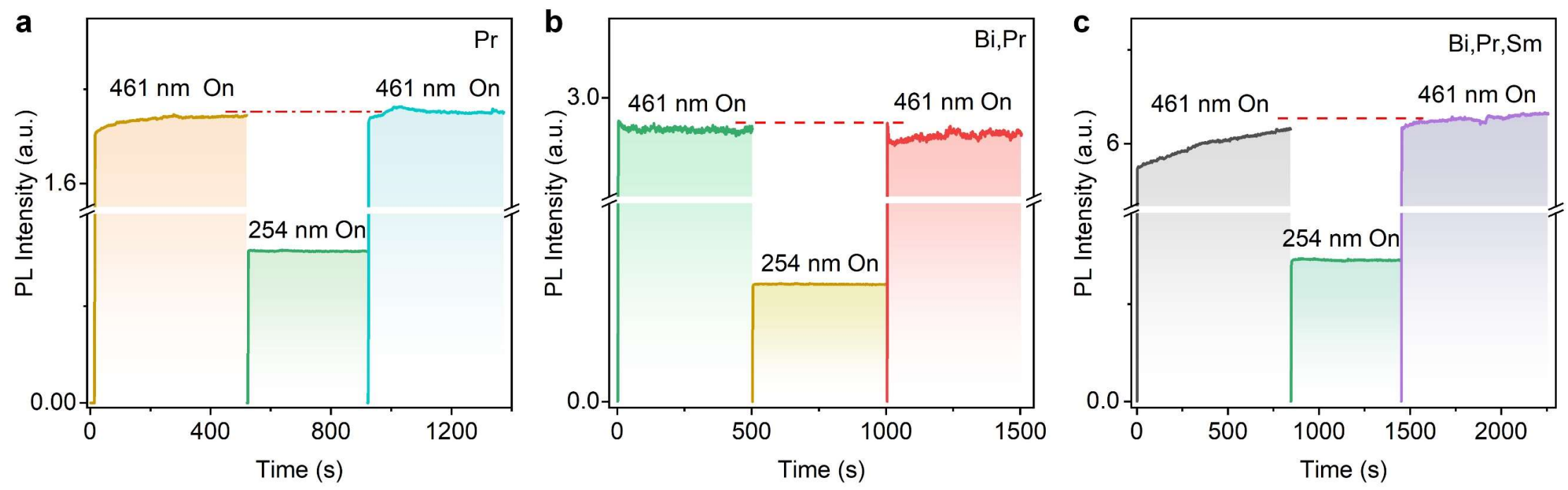

**Fig. 5.** Dynamic PL process monitoring at 620 nm upon irradiation of 254 nm and 461 nm light before and after charging in LNO:Pr, LNO:Bi,Pr, and LNO:Bi,Pr,Sm phosphors. (a) LNO:Pr. (b) LNO:Bi,Pr. (c) LNO:Bi,Pr,Sm. Note that all samples are charged using 254 nm UV light.

To further understand the roles of various dopants and the formation of traps, the bandgap structure, partial density of states (PDOS) and the charge density difference for $Bi^{3+}$ or $Pr^{3+}$ doped LNO matrix lattice are shown in Fig. 6. Electronic structural characteristics of the pure LNO that belongs to the indirect band-gap material with a bandwidth of 3.57 eV are shown in Fig. S6a for the convenience of comparison. The PDOS analysis shows that its top of valence band (VB) is mainly composed of O-2p and Nb-4d, and the bottom of CB is mainly assigned to O-2p (Fig. S6b). It is very important to note the presence of new impurity levels composed of Bi-6s and O-2p (Fig. 6a,b,d,e) near the bottom/top of CB/VB after $Bi^{3+}$ occupies $Li^{+}$/$Nb^{5+}$ site into LNO lattice. The result indicates that $Bi^{3+}$ can serve as both an electron trap and a hole trap, depending on the occupied lattice sites. While $Pr^{3+}$ replaces $Li^{+}$ ions, new impurity bands composed of Pr-4f and Nb-4d appear at the bottom of CB (Fig. 6g,h). Since this impurity band is almost at the bottom of CB, it serves as a charge transfer band between Pr and Nb instead of a stable trap energy level in the afterglow process. Compared to bandgap of the undoped LNO (Fig. S6a), the bandgap structure width decreases from 3.57 to 3.31 eV (Fig. 6d,g) when $Nb^{5+}/Li^{+}$ ions are replaced by $Bi^{3+}$ or $Pr^{3+}$. If the $Li^{+}$ site is occupied by $Bi^{3+}$, the bandgap width remains almost unchanged (Fig. 6a).

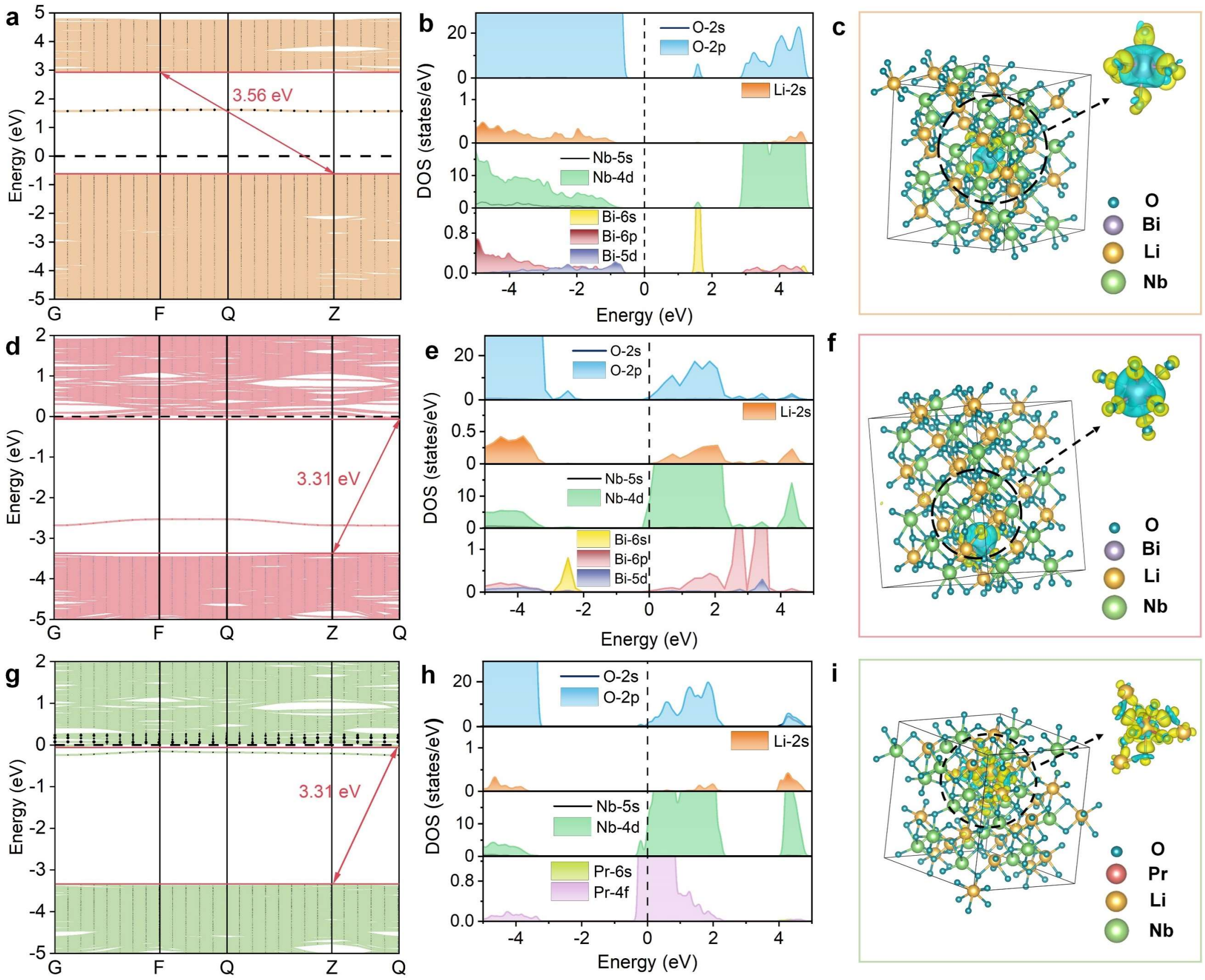

**Fig. 6.** The band-structure, PDOS and charge density difference of LNO:Bi and LNO:Pr phosphors. (a-c) LNO:Bi, where Bi substituted for Li (record as $Bi_{Li}$). (d-f) LNO:Bi, where Bi substituted for Nb (record as $Bi_{Nb}$). (g-i) LNO:Pr, in which the Pr substituted for Li (record as $Pr_{Li}$). Yellow and blue stand for the regions of electron accumulation and depletion, respectively.

Considering that in Fig. 2e, almost the same TL curve morphology is displayed in the three single doped or co-doped samples, except for different TSL intensity. The DFT calculation results indicate that $Pr^{3+}$ doping introduces charge-transfer band between Pr-4f and Nb-4d, while $Bi^{3+}$ doping introduces trap energy levels assigned to a hybrid orbital of Bi-6s and O-2p. This explains well that in the three samples, the trap types originate from the same matrix related intrinsic defects, but the trap density is different in three samples due to stabilizing and enhancing trap density by Bi/Sm doping via coupling between Bi-6s and O-2p. The DFT calculation results also provides excellent support for dynamic PL analysis of $Pr^{3+}$ (Fig. 5).

One can clearly observe changes in the charge distribution around doped atoms (Fig. 6c,f,i) based on the different electronegativity of Bi/Pr and O atoms. With Bi doping, there is charge depletion near Bi atoms, while adjacent O atoms exhibit a charge accumulation, indicating a significant charge transfer from Bi atoms to O atoms (Fig. 6c,f). This significant charge transfer subsequently leads to the formation of ionic bonds between Bi and O, which is consistent with the generation of Bi-O impurity energy levels in the bandgap (Fig. 6a,d).

Upon closer observation, it can be found that when Bi occupies $Nb^{5+}$ sites, nearby O atoms also have localized electron depletion (Fig. 6f), suggesting that Bi as electron traps may capture electrons from O-2p of CB (Fig. 6d,f). Careful observation reveals the presence of the accumulation of electron clouds between Pr and O (Fig. 6i), suggesting a strong covalent bond. Therefore, it is difficult for $Pr^{3+}$ to capture electrons from the CB for storage, indicating that $Pr^{3+}$ prefers to be a luminescent center rather than a trap center. As a result, the DFT calculation indicate that doping $Bi^{3+}$ lead to hybrid energy level of Bi-O as traps, while doping $Pr^{3+}$ acts as the luminescent center. This also explains why we observed fluorescence emission originating from $Pr^{3+}$, even interstitial oxygen emission, but did not observe fluorescence emission from $Bi^{3+}$ in LNO:Bi,Pr and LNO:Bi,Pr,Sm phosphors. While even in $Bi^{3+}$ single doped samples, there was only extremely weak blue light emission from $Bi^{3+}$ (Fig. 2a).

It is reported that the intrinsic defects of LNO are usually dominated by lithium vacancies ($V_{Li}'$) [15,56-58] due to the high volatility of lithium during high-temperature solid-state reactions. Oxygen vacancies are common defects in oxides. Considering the Bi/Pr-dopant occupied sites in LNO, there are the two and one possible substitute modes for $Bi^{3+}$ and $Pr^{3+}$, respectively, that can be expressed as:

$$\mathrm{Bi + LNO = LNO{:}Bi_{Li}^{\bullet\bullet} + 2V_{Li}' + 3Li} \tag{2}$$

$$\mathrm{Bi + LNO = LNO{:}Bi_{Nb}'' + V_{O}^{\bullet\bullet} + Nb} \tag{3}$$

$$\mathrm{Bi + LNO = LNO{:}Bi_{Nb}'' + 2V_{O}^{\bullet} + Nb} \tag{4}$$

$$\mathrm{Pr + LNO = LNO{:}Pr_{Li}^{\bullet\bullet} + 2V_{Li}' + 3Li} \tag{5}$$

It is reported that $Sm^{3+}$ exists as an electron trap in $BaZrSi_3O_9{:}Eu^{2+}$, $Sm^{3+}$ for the afterglow [59]. The doping of trace $Sm^{3+}$ occupied $Li^{+}$ site in present LNO:Bi,Pr system, disturbing the lattice for facilitating the doping of the first $Bi^{3+}$, and then opening up the Pandora's box effect of $Bi^{3+}$ doping. Therefore, in LNO:Bi,Pr,Sm sample, there may be four types of hybrid defect strings including

$Bi_{Li}^{\bullet\bullet}$ + $2V_{Li}'$, $Bi_{Nb}''$ + $2V_O^{\bullet}$, $Bi_{Nb}''$ + $V_O^{\bullet\bullet}$ and $Pr_{Li}^{\bullet\bullet}$ + $2V_{Li}'$. Combining DFT calculations, TL curves, and spectroscopic analysis, $Pr^{3+}$ doping mainly serves as the luminescent center, while $Bi^{3+}$ serves as the trap center (Fig. 6). We replaced $Li^+$ with $Na^+$ in the LNO-based series samples and found that the PL, afterglow and TSL intensity in $NaNbO_3$-based series samples would greatly reduce (Fig. S7 and S8), which indirectly supports the importance of $Li^+$. According to the vacuum ionization energy (the binding energies are 533.7 and 532.3 kJ·mol$^{-1}$ for $V_O^{\bullet}$ and $V_O^{\bullet\bullet}$, respectively) [60,61]. trap I, trap II and trap III on TL curves in LNO:Bi,Pr,Sm sample can be assigned to $Bi_{Nb}''$ + $V_O^{\bullet\bullet}$, $Bi_{Nb}''$ + $2V_O^{\bullet}$ and $Bi_{Li}^{\bullet\bullet}$ + $2V_{Li}'$ defect strings, respectively.

*3.4 PersL Mechanism*

Based on the TL curve and spectral characteristics, the possible PL, PSL and TL mechanisms of the prepared afterglow phosphor are proposed in the Fig. 7. Under UV excitation, photo generated electrons are elevated to O-2p/Nb-4d orbitals of CB/excited state from VB/$^3H_4$ of $Pr^{3+}$, and then may be trapped by $V_O^{\bullet\bullet}$ or directly return to the excited state levels of $Pr^{3+}$, leading to PL assigned to $^1D_2 \rightarrow {}^3H_4$ transition. Simultaneously, photo generated holes are captured by $V_{Li}'$ [60]. Under thermal/photo-stimulation, trapped electrons can obtain sufficient energy to escape from the trap center, and electrons/holes can be combined with $Pr^{3+}$, leading to PersL/PSL/TSL from $Pr^{3+}$. It is worth noting that in LNO based phosphors, the energy storage traps are mainly $V_O^{\bullet\bullet}$ and $V_{Li}'$ pairs. Before and after charging, there is no significant gaining or loss of electrons in $Pr^{3+}$. Consequently, we conclude that the ground state electrons of $Pr^{3+}$ are excited to the charge transfer band (CTB) and then thermally ionized to CB. Finally, they are captured by electron traps. At the same time, VB electrons are also thermally excited to the ground state $^3H_4$ of $Pr^{3+}$ and hole is trapped by hole traps, which keeps the valence of $Pr^{3+}$ unchanged. In the afterglow process, the transfer of afterglow energy may also occur from $Pr^{3+}$ to interstitial oxygen atoms, resulting in weak red afterglow emission at 716 nm [54]. The mainly reversible redox equation by photo/thermo-induced electron transfer is as follow:

$$\text{Anode reaction: } Bi^{3+} + V_O^{\bullet\bullet} + 3e \underset{kT/NIR}{\overset{UV}{\rightleftharpoons}} Bi^{2+} + V_O \quad (6)$$

$$\text{Cathodic reaction: } Bi^{3+} + 2V_{Li}' - 3e \underset{kT/NIR}{\overset{UV}{\rightleftharpoons}} Bi^{4+} + 2V_{Li} \quad (7)$$

$$\text{Total reaction: } 2Bi^{3+} + V_O^{\bullet\bullet} + 2V_{Li}' \underset{kT/NIR}{\overset{UV}{\rightleftharpoons}} Bi^{2+} + V_O + Bi^{4+} + 2V_{Li} \quad (8)$$

The entire charging to discharging process is like a rechargeable primary battery process. Under illumination of UV light, the electron of VB is elevated to CB, followed by the supply using electronics of $Bi^{3+}$ + $2V_{Li}'$ to VB. The electrons are transported through CB and then captured by the $Bi^{3+}$ octahedron with a $V_O^{\bullet\bullet}$ (Equations 6 and 7, corresponds to the anodic reaction of the electrolytic cell), completing the energy storage process by converting light energy into chemical energy. Finally, under thermal or light stimulation, the $Bi^{2+}$ + $V_O^{\bullet\bullet}$ releases trapped electrons to CB, and then these electrons are captured by the luminescent center $Pr^{3+}$, leading to radiative transition. The role of VB and CB is similar to that of strong electrolyte solutions, helping to form closed circuits. As shown in Fig. 7, path 1 represents the oxidation process of $Bi^{3+}$ + $V_O^{\bullet\bullet}$ under UV irradiation and the reduction process of $Pr^{3+}$ + $V_{Li}'$, which is the energy storage process of converting UV light energy into chemical energy. Path 2 represents the process in which chemical energy is converted into light energy through the binding of electron-hole pairs by $Pr^{3+}$ under photo/thermo perturbation.

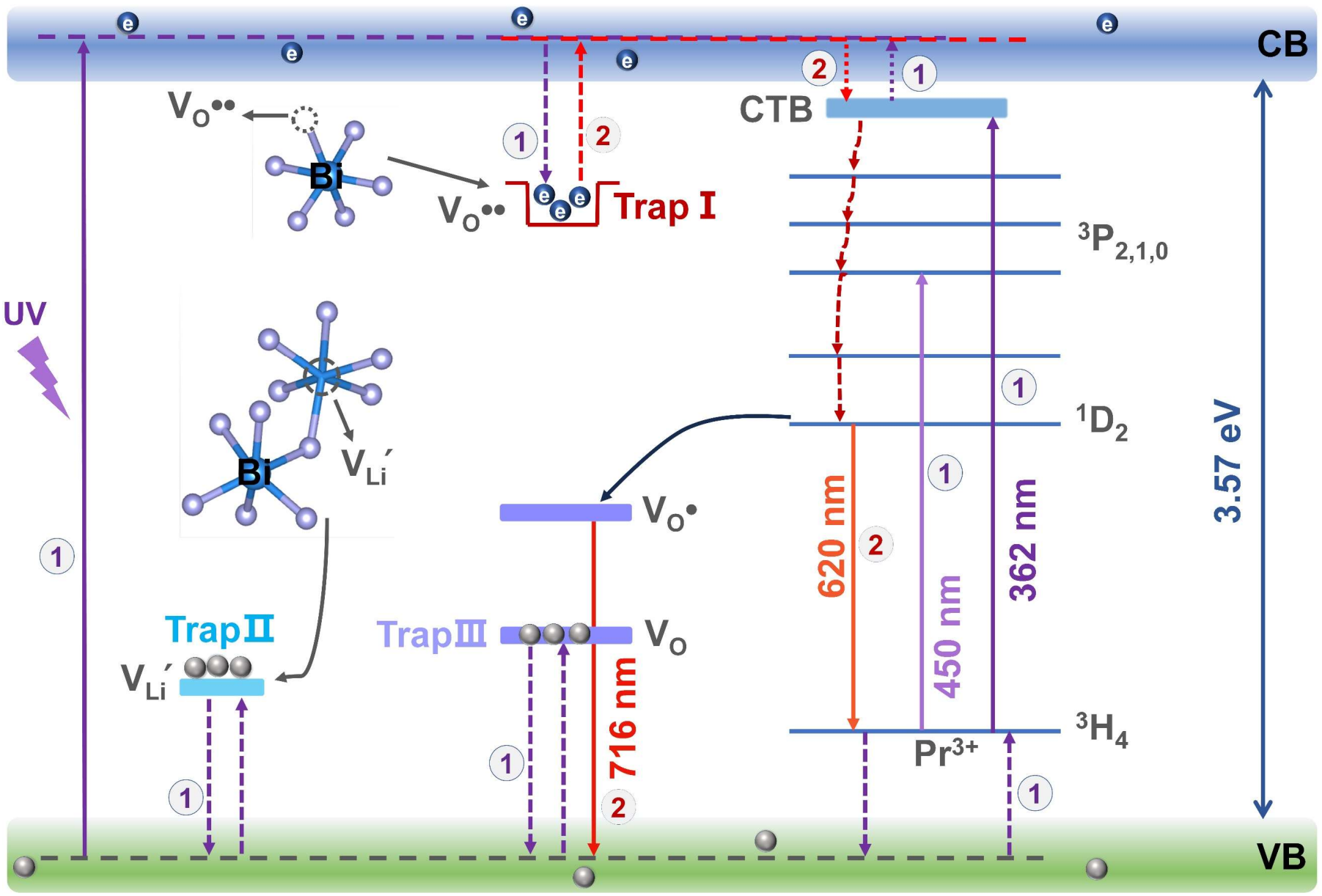

**Fig. 7.** Simple schematic diagram of mechanisms for PL, PersL, TL, PSL, and PSPL in the as-synthesized LNO:Bi,Pr,Sm/LNO:Bi,Pr samples. The corresponding three types of defect strings have been drawn and marked as traps in the Figure. Note that purple path 1 represents the electron flow during the charging process, and red path 2 represents the electron flow during the luminescence process.

*2.5. Optical Information Storage and Intelligent Anti-Counterfeiting*

The excellent optical performances of the LNO:$Bi^{3+}$ and LNO:$Bi^{3+}$,$Pr^{3+}$ sample allow them to serve as inks to design and fabricate charming multi-color dynamic patterns like chameleon, as shown in Fig. 8. Fig. 8a is a structural diagram of “pomegranate flowers” pattern, where the “pomegranate flowers” and “leaf” are printed by using LNO:Bi,Pr,Sm and LNO:Bi phosphors, respectively, on non-fluorescent paper. As the irradiation light wavelength changes from 254 nm via 365 to 488 nm, the PL color of leaves is changeable from blue via cyan to green, while PL color of flowers ranges from rose red to red to peach red (Fig. 8b-d). When the irradiation light source is turned off, the 365 nm pre-irradiated pattern displays the optimum bright red PersL flower patterns with readable green luminescence leaves (Fig. 8c), while the 254/488 nm pre-irradiated patterns only displays red PersL flowers without identifiable leaves (Fig. 8b,d). The disappearance of the afterglow leaves takes about 30 seconds after turning off the light source (Fig. 8c), while the afterglow flowers can last for 2 min (Fig. 8c,d). The disappeared afterglow flowers can be illuminated again through NIR laser or heating, while the leaves are opposite (Fig. 8e,f). Interestingly, TSL intelligently responds to different local temperatures with varying brightness levels, forming contrast images that combine visual temperature sensing with anti-counterfeiting integration. Specifically, luminescence intelligently remembers the temperature of the charging area at that time through brightness response. For example, in the TSL photo in Fig. 8f, information is extracted under the same thermal excitation. The brightest area represents the charging temperature of 85 °C at that time, the second brightest area represents the room temperature charging area, and the darker area represents the charging temperature at 200 °C. These TSL photos integrate temperature memory and reconnaissance functions for encryption and charging environment.

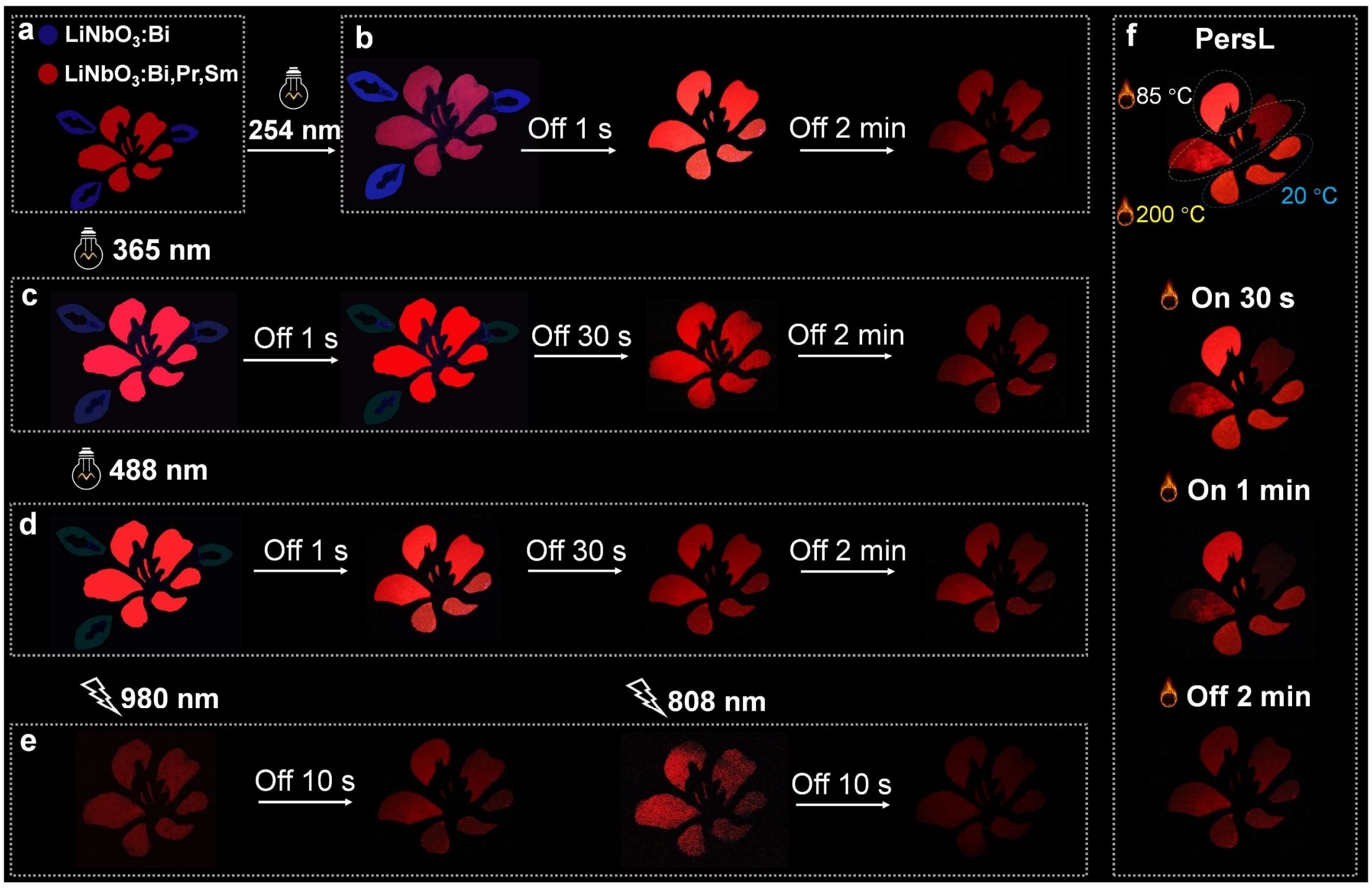

**Fig. 8.** Multi-modal images of a "pomegranate flowers" pattern printed using LNO:$Bi^{3+}$ and LNO:Bi,Pr,Sm phosphors for information storage and encryption. (a) Structural design of "pomegranate flowers" pattern. (b) PL and PersL images under and after 254 nm UV light irradiation. (c) PL and PersL images upon and after 365 nm UV light irradiation. (d) PL and PersL images under and after 488 nm UV light irradiation. (e) PSL and PSPL images upon and after irradiation of a 980/808 nm NIR laser (0.6 W). (f) Temperature intelligent response of PSL contrast photographs after pre-irradiation of 365 nm.

The remote intelligent encryption with a numeric format is implemented on the memory phosphor film. The UV pattern mask was fabricated using a lithography machine, and a fixed infrared laser as shown in Fig. 9a for disk-shaped digital data storage. The information recording layer is prepared by coating the discs with LNO:Bi,Pr,Sm and PMMA. The unit of light emission is represented as 1, while the unit of no emission is recorded as 0, as shown in Fig. 9b. Multi-level data are written into traps of different depths in the luminescent disk through temperature management in Fig. 9b. Specifically, the first level data sequence of 14 bits (00100101000100) is written into luminescent film by Tom with patterning 365 nm UV light mask at 473 K in location A. Then, he wrote the second sequence (00100101000000) at 393 K, third sequence (00100100000000) at 353 K, and fourth data sequence (00100000000000) at 293 K at the same position on the disc. Mary in location B needs to extract the correct temperature command in order to obtain the correct four layers encrypted information. According to the decryption instructions given by Tom, the first layer signal (at a temperature of 293~313 K), the second layer encrypted signal (at a temperature of 353~373 K), the third layer signal (at a temperature of 393~413 K), and the fourth layer signal (at a temperature of 473~493 K) can be read in the correct temperature sequence by using a camera for capturing photos. Therefore, the above encryption and decryption processes are not only related to the TSL characteristics of the phosphor, but also to the writer's subjective information in the trap dimension, improving the security level to a new level.

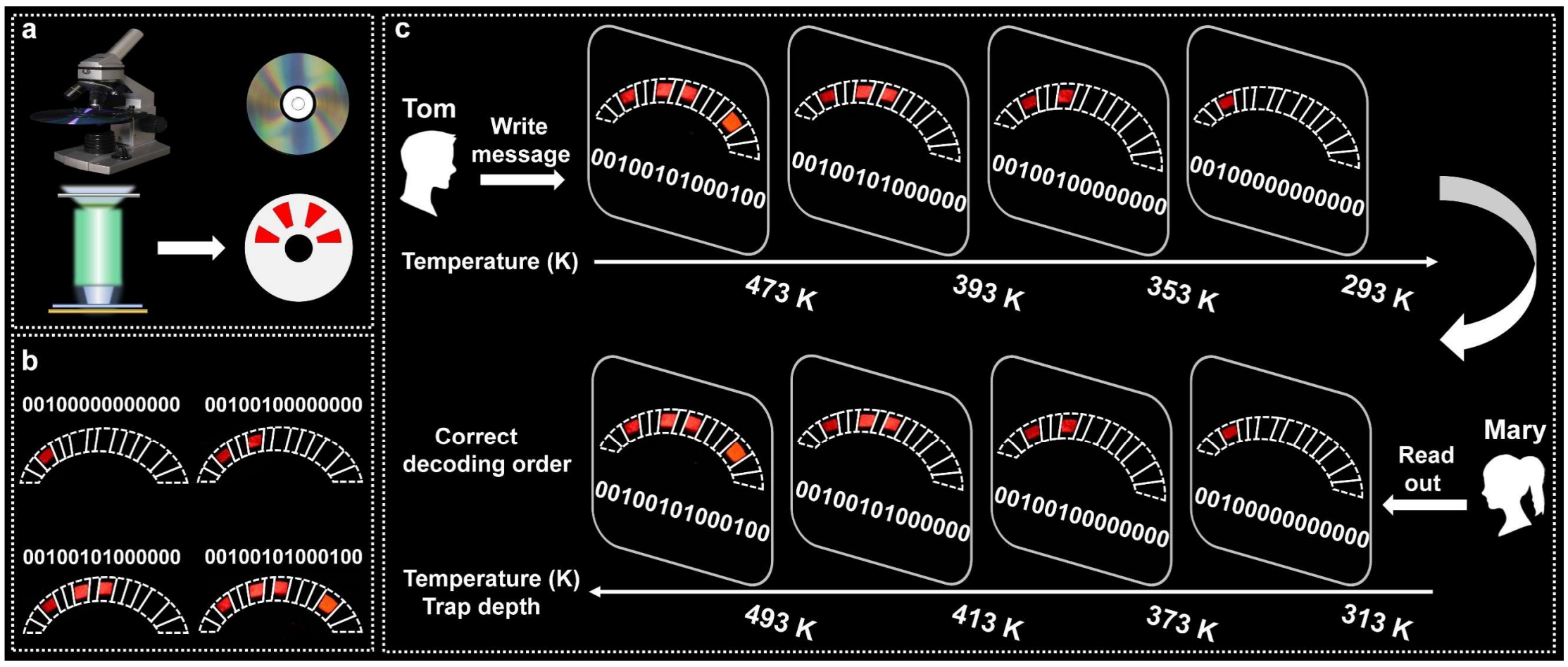

**Fig. 9.** Multi-level data encryption using memory storage LNO:Bi,Pr,Sm phosphor film by managing temperature for remote intelligent anti-counterfeiting. (a) Schematic diagram of making the optical disc, the process of decoding disc information, and the optical storage of disc. (b) The data sequence of 14 bit corresponding to pattern. (c) Remote information writing with the patterning 365 nm UV light mask and reading through temperature sequence management. Note that the temperature for writing and reading multi-level information (i.e. the depth of the access trap) has been marked on the graph.

## 4. Conclusions

In conclusion, we have successfully prepared LNO:$Bi^{3+}$,$Pr^{3+}$,$Sm^{3+}$ phosphors with enhanced and controllable TSL for remote intelligent anti-counterfeiting. The distribution and type of trap, the capture and transport processes of electrons/holes have been comprehensively surveyed through various PL spectra, PL dynamic processes, and TL technique, thermo/photon-stimulated spectra, and DFT calculations, indicating that the trap state (i.e., depth, density and type) and carrier transport process can be deliberately manipulated via doping $Bi^{3+}$ and $Sm^{3+}$ ions in LNO:Pr phosphors. PersL tuning mechanisms are proposed based on enriching trap density and a carrier transport path is proposed. Consequently, we have achieved TSL that can vary as a function of time and temperature. Furthermore, these phosphors as inks have demonstrated remote intelligent anti-counterfeiting. These findings provide a universal approach for the design of memory phosphors with multi-mode luminescence, opening up a new avenue for the emerging applications of phosphors in anti-counterfeiting and artificial intelligence temperature sensor.

**Acknowledgements**

This work was supported by the National Natural Science Foundation of China (11604253 and 51672208), Shaanxi Fundamental Science Research Project for Mathematics and Physics (23JSY003), and Shaanxi Key Science and Technology Innovation Team Project (2022TD-34).

**Conflicts of interest**

There are no conflicts to declare.

**Notes and references**